\documentclass[aps,pre,preprint,groupedaddress]{revtex4-1}

\usepackage{hyperref}
\usepackage{graphicx}
\usepackage{bm}
\usepackage{color}

\begin{document}


\title{Statistical Physics approach to dendritic computation: the
  excitable-wave mean-field approximation}




\author{Leonardo L. Gollo}%
\email{leonardo@ifisc.uib-csic.es}
\affiliation{
IFISC (CSIC - UIB), Instituto de F\'{\i}sica Interdisciplinar y Sistemas 
Complejos, Campus Universitat Illes Balears, E-07122 Palma de Mallorca, 
Spain
}%

\author{Osame Kinouchi}%
 \email{osame@ffclrp.usp.br}
\affiliation{
Laborat\'orio de F{\'{\i}}sica Estat{\'{\i}}stica e Biologia
Computacional, Departamento de F{\'{\i}}sica, FFCLRP,
Universidade de S\~ao Paulo, Avenida dos Bandeirantes 3900, 14040-901,
Ribeir\~ao Preto, SP, Brazil}%

\author{Mauro Copelli}%
\email{mcopelli@df.ufpe.br} 
\affiliation{Departamento de F{\'\i}sica,
  Universidade Federal de Pernambuco, 50670-901 Recife, PE, Brazil}%



\date{\today}

\begin{abstract}
  We analytically study the input-output properties of a neuron whose
  active dendritic tree, modeled as a Cayley tree of excitable
  elements, is subjected to Poisson stimulus. Both single-site and
  two-site mean-field approximations incorrectly predict a
  non-equilibrium phase transition which is not allowed in the
  model. We propose an excitable-wave mean-field approximation which
  shows good agreement with previously published simulation results
  [Gollo \textit{et al.}, PLoS Comput. Biol. {\bf 5}(6) e1000402 (2009)] and
  accounts for finite-size effects.  We also discuss the relevance of
  our results to experiments in neuroscience, emphasizing the role of
  active dendrites in the enhancement of dynamic range and in gain
  control modulation.
\end{abstract}


\maketitle

\tableofcontents

\section{Introduction}


Computational neuroscience is a growing field of research which
attempts to incorporate increasingly detailed aspects of neuronal
dynamics in computational models~\citep{Bower,Neuron}. Since the
pioneering work of Hodgkin and Huxley (HH)~\citep{HH52}, which unveiled
how the action potential in the giant squid axon could be described by
ordinary differential equations governing the gating of ionic
conductances across a membrane patch, the computational modeling of
neuronal biophysical processes has been done at several levels, from
whole neural networks to dendritic spines and even single ionic channel
dynamics~\citep{Dayan2001}.


Rall was probably the first to extend conductance-based modeling to
dendrites~\citep{Rall64}, starting what is nowadays a field of its own:
the investigation of so-called dendritic
computation~\citep{Stuart08}. The main theoretical tool in this
enterprise has been cable theory, the extension [via partial
differential equations (PDEs)] of the HH formalism to extended
systems, which allows one to include spatial information about
dendrites such as the variation of channel densities along the trees,
different branching patterns etc.~\citep{Koch}. The assumption that
dendrites are passive elements renders cable theory linear, allowing
the application of standard techniques from linear PDEs and yielding
insightful analytical results~\citep{Koch}. This assumption, however,
has been gradually revised since the first experimental evidences that
dendrites have nonlinear properties~\citep{Eccles58}.  A variety of
channels with regenerative properties are now identified which can
sustain the propagation of nonlinear pulses along the trees (called
{\em dendritic spikes\/}), whose functional role has nonetheless
remained elusive~\citep{Stuart08}.

The conditions for the generation and propagation of dendritic nonlinear
excitations have been investigated via cable
theory~\citep{Mel93,Stuart08} at the level of a dendritic
branchlet. This has proven useful for understanding the specific role
of each ionic channel in the dynamical properties of the nonlinear
propagation, specially in comparison with experiments, which have
mostly been restricted to the injection of current at some point in the
neuron (say, a distal dendrite) and the measurement of the membrane
potential at another point (say, the
soma)~\citep{Johnston08,Sjostrom08}. While this limitation is justified
by the difficulties of injecting currents and measuring membrane
potentials in more than a couple of points in the same neuron, we must
remember that neurons \textit{in vivo} are subjected to a different stimulus
regime, with many synaptic inputs arriving with a high degree of
stochasticity and generating several dendritic spikes which may
propagate and interact.

In this more realistic and highly nonlinear scenario, cable theory,
though still having the merit of being able to incorporate as many
ionic channels as experiments reveal, becomes analytically
untreatable. Being able to reproduce the fine-grained experimental results
of a complex system such as a neuron does not imply that the essential
aspects of its dynamics will be identified. Or, to put it in a
renormalization group parlance, ``realistic biophysical modeling''
does not allow us to separate the relevant observables from the
irrelevant ones that can be eliminated without significantly
changing some robust property of the system. In fact, this has been
recognized in the neuroscience literature, which has emphasized the
need for theoretical support~\citep{Reyes01,Herz06,London05} and witnessed the
increase of theoretical papers in the field of dendritic
computation~\citep{PoiraziMel2001,PoiraziMel03a,PoiraziMel03b,Morita09,Coop10}.

In this context, we have recently attempted to understand the behavior
of an active dendritic tree by modeling it as a large network of
interacting nonlinear branchlets under spatio-temporal stochastic
synaptic input and allowing for the interaction of dendritic
spikes~\citep{Gollo09}. With a statistical physics perspective in mind,
we have tried to incorporate in the model of each branchlet only those
features that seemed most relevant, and have investigated the
resulting collective behavior. Thus each excitable
branchlet was modeled as a simple 3-state cellular automaton, with
the propagation of dendritic spikes occurring with probabilities which
 depend on direction (to account for the differences
between forward- and backward-propagating spikes).

This model has revealed that such a tree performs a highly nonlinear
``computation'', being able to compress several decades of input rate
intensity into a single decade of output rate intensity. This signal
compression property, or enhancement of dynamic range, is a general
property of excitable media and has proven very robust against
variations in the topology of the medium and the level of modeling,
from cellular automata to compartmental conductance-based
models~\citep{Copelli02,Copelli05a,Copelli05b,Furtado06, Kinouchi06a,
  Copelli07, Wu07, Assis08, Ribeiro08a, Publio09,
  Larremore11,Larremore11b, Buckley11}. Furthermore, the idea that dynamic range can be
enhanced in neuronal excitable media has received support from
experiments in very different setups~\citep{Kihara09,Shew09}, which
again suggests that the phenomenon is robust.

Our aim here is to analytically explore the model introduced in
Ref.~\citep{Gollo09} and described in section~\ref{model}.  In
section~\ref{EWMFA} we show that the traditional
cluster approximations applied to the system master equations fail to
qualitatively reproduce the essential features observed in the simulations and
experimental data. We propose a mean-field approximation which
circumvents the problems faced by the traditional approach, yielding
good agreement with simulations. We conclude in
section~\ref{conclusions} with a discussion of the consequences of our
results for neuroscience and the perspectives for future work.

\section{Modeling an active dendritic tree}
\label{model}

\begin{figure}[!ht]
\begin{center}
\includegraphics[angle=0,width=0.98\columnwidth]{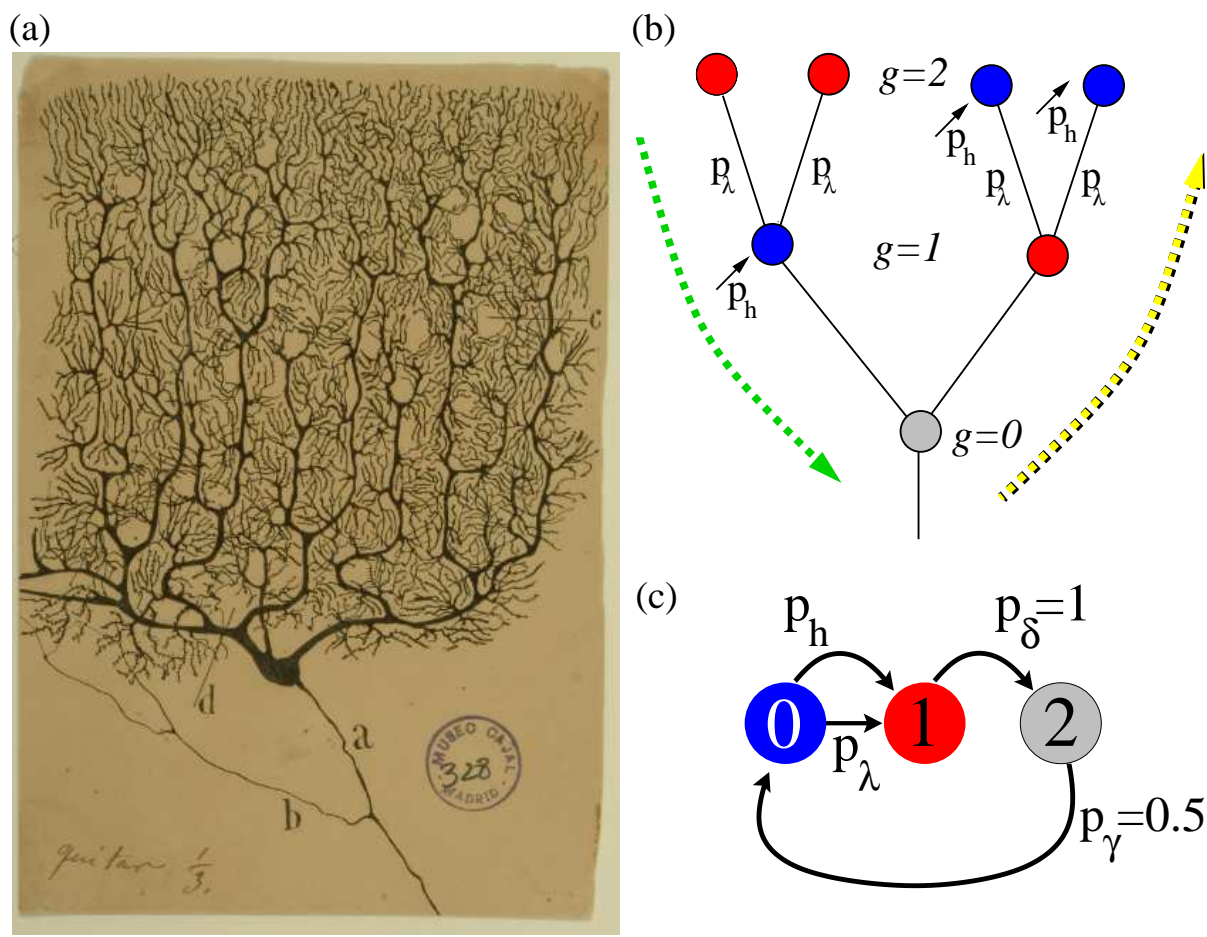}
\caption{\label{fig:dendrite} (Color online) Model of an active
  dendritic tree.  (a) A famous drawing by Ramon y Cajal of a human
  Purkinje cell.  (b) Excitable elements (circles) connected (bars) in
  a Cayley tree topology with $G=2$ layers and coordination number
  $z=3$ (one mother and $k=2$ daughter branches).  Dendritic
  branchlets are driven by independent Poisson stimuli (small
  arrows). (c) Each dendritic branchlet can be in one of three states:
  quiescent (0), active (1) or refractory (2).}
\end{center}
\end{figure}

The dendritic tree of an isolated neuron contains no loops and divides
in two daughter branches at branching points. For instance,
Fig.~\ref{fig:dendrite} (a) depicts one of Ramon y Cajal's drawings of
a human Purkinje cell, which shows a huge ramification.  Measured by
the average number $G$ of generations (i.e., the number of
branch-doubling iterations the primary dendrite undergoes), the size
of the dendritic trees can vary widely.  One can think of an active
dendritic tree as an excitable medium~\citep{Lindner04}, in which each
site represents, for instance, a branching point or a dendritic
branchlet connected with two similar sites from a higher generation
and one site from a lower generation. Correspondingly, the standard
model in this paper is a Cayley tree with coordination number
$z=3$~\citep{Gollo09}. Each site at generation $g$ has a mother branch
from generation $g-1$ and generates two daughter branches ($k\equiv
z-1 = 2$) at generation $g+1$. The single site at $g=0$ would
correspond to the primary (apical) dendrite which connects with the
neuron soma [see Fig.~\ref{fig:dendrite}(b)]. Naturally, the Cayley
tree topology of our model is a crude simplification of a real tree,
as attested by the differences between Figs.~\ref{fig:dendrite} (a) and~\ref{fig:dendrite}(b).

Each site represents a dendritic branchlet, which we model with a
three-state excitable element~\citep{Lindner04}: $x_i(t)\in\{0,1,2\}$
denotes the state of site $i$ at time $t$. If the branchlet is active
($x_i=1$), in the next time step it becomes refractory ($x_i=2$) with
probability $p_\delta$. Refractoriness is governed by $p_\gamma$,
which is the probability with which sites become quiescent ($x_i=0$)
again [see Fig.~\ref{fig:dendrite}(c)].  Here we have used
$p_\delta=1$ and $p_\gamma=0.5$. The propagation of dendritic spikes
along the tree is assumed to be stochastic as well: each active
daughter branch can independently excite its mother branch with
probability $p_\lambda$, contributing to what is referred to as
forward propagation [i.e., from distal dendrites to the soma, see large
descending arrow in Fig.~\ref{fig:dendrite}(b)]. Backpropagating
activity is also allowed in the model, with a mother branch
independently exciting each of its quiescent daughter branches with
probability $\beta p_\lambda$ [large ascending arrow in
Fig.~\ref{fig:dendrite}(b)], where $0\leq \beta \leq 1$.


Dendrites are usually regarded as the ``entry door'' of information
for the neuron, i.e., the dominant location where (incoming) synaptic
contacts occur. Our aim then is to understand the response properties
of this tree-like excitable medium. Incoming stimulus is modeled as a
Poisson process: besides transmission from active neighbors (governed
by $p_\lambda$ and $\beta$), each quiescent site can independently
become active with probability $p_h \equiv 1-\exp(-h\Delta t)$ per
time step [see Fig.~\ref{fig:dendrite}(c)], where $\Delta t=1$~ms is an
arbitrary time step and $h$ is referred to as the stimulus
intensity. It reflects the average rate at which branchlets get
excited, after the integration of postsynaptic potentials, both excitatory
and inhibitory~\citep{Gollo09}. With synchronous update, the model is
therefore a cyclic probabilistic cellular automaton.

A variant of the model accounts for the heterogeneous distribution of
synaptic buttons along the proximal-distal axis in the dendritic
tree. It consists of a layer-dependent rate $h(g)=h_0 e^{a g}$, with
$a$ controlling the nonlinearity of the dependence~\citep{Gollo09}. We
will mostly restrict ourselves to the simpler cases $\beta=1$ and
$a=0$.

\subsection{Simulations}

In the simulations, the activity $F$ of the apical ($g=0$) dendritic
branchlet is determined by the average of its active state over a
large time window ($T=10^4$ time steps and 5 realizations). The
response function $F(h)$ is the fundamental input-output neuronal
transformation in a rate-code scenario (i.e., assuming that the mean
incoming stimulus rate and mean output rate carry most of the
information the neuron has to transmit).  

In a never ending matter of investigation, rate code has historically
competed with temporal code, which is also supported by plenty of
evidence~\citep{Rieke97}. Auditory coincidence
detection~\citep{Agmon-Snir98}, as well as spatial
localization properties of place and grid cells fundamentally depend on the precise spike
time~\citep{Moser08}. Spike-timing-dependent plasticity, responsible
for memory formation and learning, critically relies on small time
differences (of order of tens of milliseconds) between presynaptic and postsynaptic
neuronal spikes~\citep{Bi98}. Moreover, zero-lag or near zero-lag
synchronization, which are thought to play an active role in cognitive
tasks~\citep{Uhlhaas09}, has been recently shown to be supported and
controlled by neuronal circuits despite long connection
delays~\citep{Fischer06,Vicente08,Gollo10,Gollo11}.  Nevertheless,
because of its robustness to the high level of stochasticity and
trial-to-trial variability present in the brain~\citep{Rolls10}, rate
code is probably more globally found~\citep{Buesing11,Rolls11}.  In
this paper we implicitly assume that rate code holds.

\begin{figure}[!ht]
\centerline{\includegraphics[angle=0,width=0.83\columnwidth]{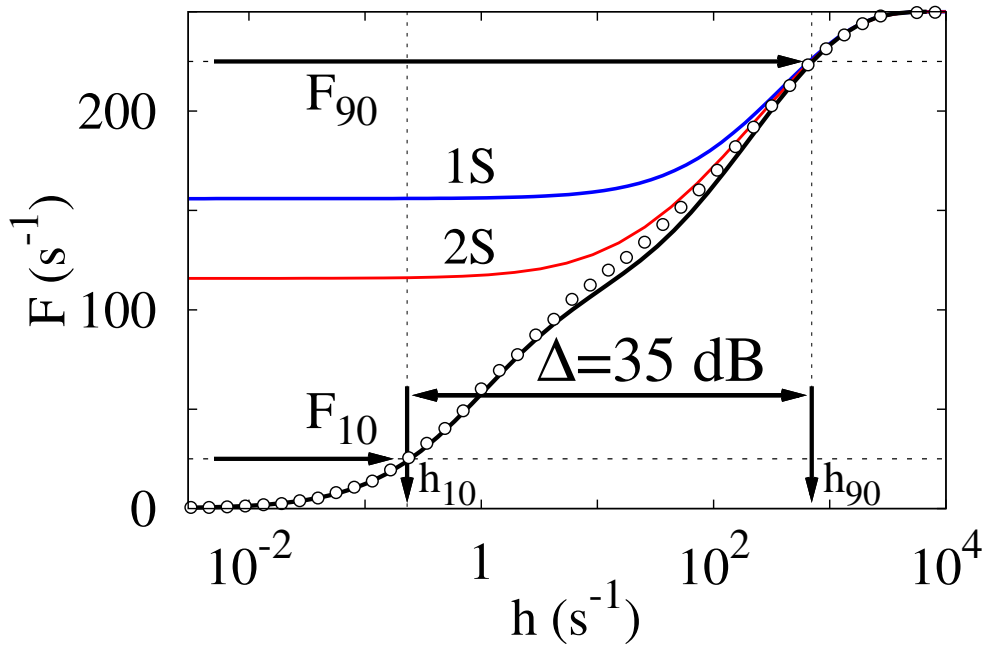}}
\caption{\label{figCRandDRdef}(Color online) Response curve $F(h)$ for
  simulations (symbols) and mean-field approximations (lines; 
  1S, 2S and black for EW) for $p_\lambda=0.7$ and
  $G=10$. Horizontal and vertical arrows show the relevant parameters
  for calculating the dynamic range $\Delta$ (see Eq.~\ref{Delta}).}
\end{figure}

A typical response curve obtained from simulations with
$p_\lambda=0.7$ and $G=10$ is shown in Fig.~\ref{figCRandDRdef}
(symbols). It is a highly nonlinear saturating curve, with the
remarkable property of compressing decades of stimulus intensity $h$
into a single decade of apical response $F$. A simple measure of this
signal compression property is the dynamic range $\Delta$, defined as
\begin{equation}
\label{Delta}
\Delta = 10\log\left(\frac{h_{90}}{h_{10}}\right)\; ,
\end{equation}
where $h_{x}\equiv F^{-1}(F_x)$ is the stimulus value for which the
response reaches $x\%$ of its maximum range: $F_x\equiv
F_{min}+\frac{x}{100}(F_{max}-F_{min})$, where $F_{min}=\lim_{h\to
  0}F(h)$ and $F_{max}=\lim_{h\to\infty}F(h)$. As exemplified in
Fig.~\ref{figCRandDRdef}, $\Delta$ amounts to the range of stimulus intensities
(measured in dB) which can be appropriately coded by $F$, discarding
stimuli which are either so weak as to be hidden by the self-sustained
activity of the system ($F<F_{10}$) or so strong that the response is, in
practice, non-invertible owing to saturation ($F>F_{90}$).

Several features of this model have been explored
previously~\citep{Gollo09}, like the dependence of $\Delta$ on model
parameters, the double-sigmoid character of the response function, as
well as the robustness of the results with respect to variants which
increase biological plausibility. All these were based on simulations
only. We now attempt to reproduce the results analytically.


\section{Mean-Field Approximations}
\label{EWMFA}

\subsection{Master equation}

The system can be formally described by a set of master equations. For
the general case of arbitrary $k$, let
$P_t^g\left(x;y;u^{(i)},v^{(j)},\ldots\right)$ be the joint
probability that at time $t$ a site at generation $g$ is in state $y$,
its mother site at generation $g-1$ is in state $x$, $i$ ($j$) of its
daughter branches at generation $g+1$ are in state $u$ ($v$) etc. 

Although the results in this paper are restricted to trees with $k=2$,
for completeness we write down the master equation for general $k$.
The explicit derivation of the master equations for any layer is shown
in Appendix~\ref{derivation}. The equations for $0<g<G$ can be written
as follows:
\begin{eqnarray}
\label{P(1)}
P_{t+1}^{g}(;1;) & = & P_{t}^{g}(;0;)  - \left(1-
p_h \right)  \sum^{k}_{i=0}\biggl[   p_{\lambda}^i {k
\choose i} (-1)^i P_t^{g}\left(;0;1^{(i)}\right) \nonumber \\
 & & \nonumber \\
 & &- \beta p^{i+1}_\lambda{k \choose i}(-1)^i
P_t^{g}\left(1;0;1^{(i)}\right)\biggr] +(1-p_\delta) P_t^{g}(;1;)\;,  \\
\label{P(2)}
P_{t+1}^{g}(;2;) & = & p_\delta P_{t}^{g}(;1;) + (1-p_\gamma)
P_{t}^{g}(;2;)\; , \\
\label{P(0)}
P_{t+1}^{g}(;0;)& = & 1 - P_{t+1}^{g}(;1;) - P_{t+1}^{g}(;2;)\;,
\end{eqnarray}
where $P_t^g(x;y;w^{(0)})\equiv P_t^g(x;y;)$ is a two-site joint
probability and $P_t^g(;y;)$ [also written $P_t^g(y)$ for simplicity]
is the probability of finding at time $t$ a site at generation $g$ in
state $y$ (regardless of its neighbors).

Equations for the central ($g=0$) and border ($g=G$) sites can be
obtained from straightforward modifications of Eq.~\ref{P(1)},
rendering

\begin{eqnarray}
\label{P0(1)}
P_{t+1}^{0}(;1;) & = & P_{t}^{0}(;0;)  - \left(1-
p_h \right)  \sum^{k+1}_{i=0}\biggl[   p_{\lambda}^i {k+1
\choose i} (-1)^i P_t^{0}\left(;0;1^{(i)}\right)\biggr] \nonumber \\
 & & +(1-p_\delta) P_t^{0}(;1;)\;,  \\
\label{PG(1)}
P_{t+1}^{G}(;1;) & = & P_{t}^{G}(;0;) + (1-p_h)\left[\beta p_\lambda P^{G}_t(1;0;)\right]   +(1-p_\delta) P_t^{G}(;1;)\;, 
\end{eqnarray}
whereas Eqs.~\ref{P(2)}~and~\ref{P(0)} remain
unchanged. Naturally, the full description of the dynamics would
require higher-order terms (infinitely many in the limit
$G\to\infty$), but Eqs~\ref{P(1)}~to~\ref{P(0)} suffice to yield
the mean-field equations we address below.

\subsection{Single-site mean-field approximation}

The simplest method for truncating the master equations is the
standard single-site (1S) mean-field approximation~\citep{Marro99},
which results from discarding the influence of any neighbors in the
conditional probabilities:  $P_t^g(y|x) \equiv
P_t^g(;y;x)/P_t^{g+1}(;x;)\stackrel{(1S)}{\approx} P_ t^g(y)$. If this
procedure is applied separately for each generation $g$, one obtains
the factorization
$P_t^g\left(x;y;u^{(i)},v^{(j)}\right)\stackrel{(1S)}{\approx}
P_t^{g-1}(x) P_t^g(y) [P_t^{g+1}(u)]^i [P_t^{g+1}(v)]^j$, which
reduces the original problem to a set of coupled equations for
single-site probabilities:
\begin{eqnarray}
\label{P1g}
P^{g}_{t+1}(1)& \stackrel{(1S)}{\simeq}& P^g_t(0) \Lambda^g(t) +
(1-p_\delta) P_t^{g}(1) \; ,
\end{eqnarray} 
where 
\begin{equation}
\label{Lambda}
\Lambda^g(t)= 1-(1-p_h)\left[1-\beta p_\lambda
  P^{g-1}_t(1)\right]\left[1-p_\lambda P^{g+1}_t(1)\right]^{k}
\end{equation}
is the probability of a quiescent site becoming excited due to either
an external stimulus or propagation from at least one of its $z=k+1$
neighbors (i.e., for
$0<g<G$). For $g=0$ and $g=G$ one has \\
\begin{eqnarray}
\Lambda^0(t)  &=& 1-(1-p_h)\left[1-p_\lambda P^{1}_t(1)\right]^{k+1} \; , \\
\label{eq:LambdaG}
\Lambda^G(t) &=& 1-(1-p_h)\left[1-\beta p_\lambda
  P^{G-1}_t(1)\right]  \; .
\end{eqnarray}
Note that this approximation retains some spatial information through
its index $g$, whereby the generation-averaged activation $P_t^g(1)$
is coupled to $P_t^{g+1}(1)$ and $P_t^{g-1}(1)$, rendering a
$2(G+1)$-dimensional map as the reduced dynamics [note that the
dimensionality of the probability vector is $3(G+1)$, but
normalization as in Eq.~\ref{P(0)} reduces it to $2(G+1)$]. Although
this facilitates the incorporation of finite-size effects (which are
necessary for comparison with finite-$G$ system simulations),
we will see below that the results are not satisfactory. In fact,
the results are essentially unchanged if we further collapse the different
generations: $P_t^g(x)=P_t(x)$, $\forall g$ (which is the usual
mean-field approximation, implying surface terms are to be neglected
in the limit $G\to\infty$). The reasons for keeping a generation
dependence will become clear when we propose a different approximation
(see section~\ref{EW}).

\begin{figure}[!ht]
\centerline{\includegraphics[angle=0,width=0.99\columnwidth]{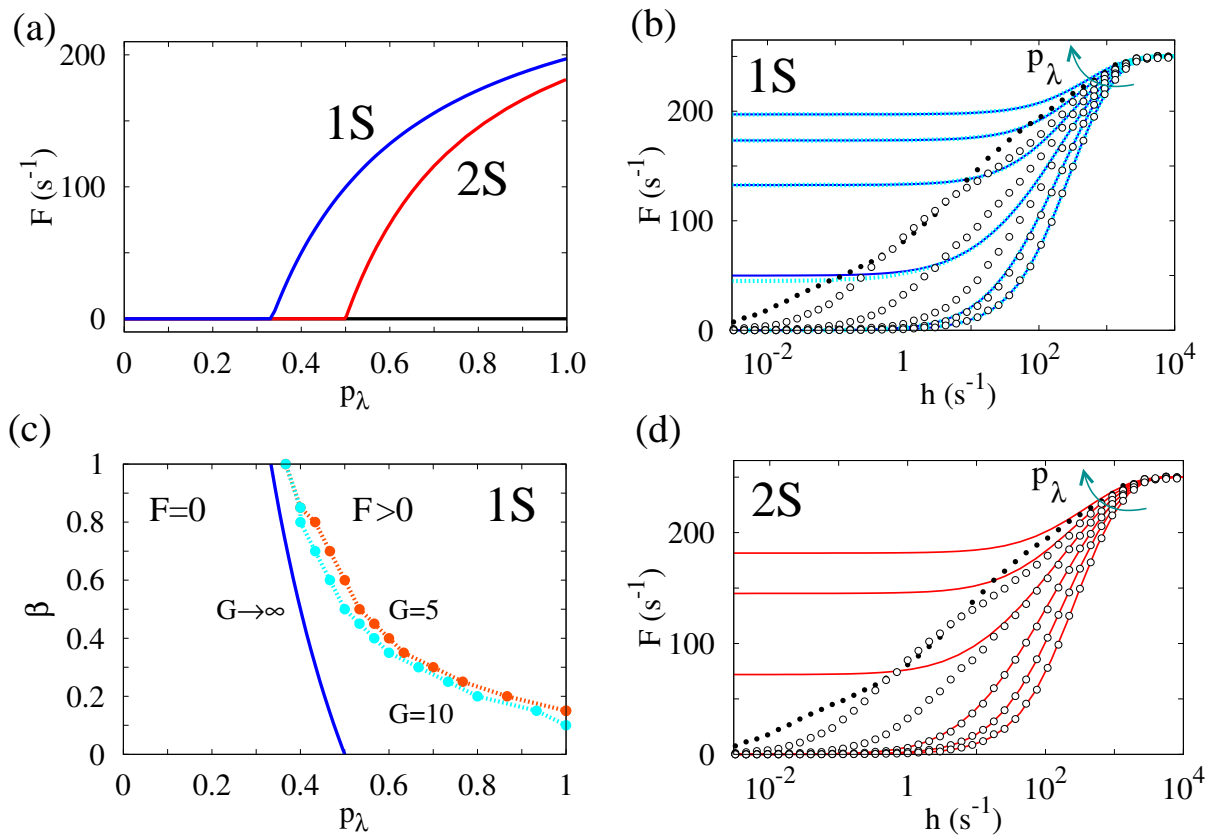}}
\caption{\label{fig1S2S}(Color online) Mean-field approximations. (a)
  Firing rate of the 1S, 2S and EW (black) approximations in the
  absence of stimulus ($h=0$). (b) Family of response functions for
  $\beta=1$ and $p_\lambda=0, 0.2,0.4,\ldots, 1$. Symbols represent
  simulations (as in Ref.~\cite{Gollo09}) and curves are the 1S
  mean-field approximation. Open (closed) symbols correspond to
  probabilistic (deterministic, $p_\lambda=1$) neighbor coupling, and
  dotted (continuous) curves correspond to $G=10$ ($G\to\infty$). (c)
  Phase diagram under the 1S approximation for different system sizes.
  (d) Same as (b) but for 2S approximation with $G\to\infty$.}
\end{figure}

To compare our results with the case of interest for real
dendrites, in the following we restrict ourselves to the binary tree,
namely $k=2$. Figure~\ref{fig1S2S}(a) shows the results for the
stationary value $F \equiv \frac{1}{\Delta t} \lim_{t\to\infty} P_t^0(1)$ in the absence
of stimulus, i.e., the fixed point of the 1S mean-field equations for
$h=0$, as a function of the branchlet coupling $p_\lambda$. The parameter
values are $G=10$, $\beta=1$ and $p_\delta=1$ (deterministic spike
duration). In the absence of stimulus, we see that the 1S
approximation predicts a phase transition at $p_\lambda = p_{\lambda
  c}^{(1S)} = 1/3$. As a consequence, the response curve $F(h)$ for
$p> p_{\lambda c}^{(1S)}$ displays a plateau in the limit $h\to 0$, as
shown in Fig.~\ref{fig1S2S}(b). The 1S approximation yields results
comparable to simulations only below $p_{\lambda c}^{(1S)}$, but
performs rather poorly above the phase transition it
predicts. However, given the deterministic spike duration (the only
state in which a given site can excite its neighbors) and the absence
of loops in the topology, a stable phase with stable self-sustained
activity cannot exist~\citep{Furtado06,Gollo09}.

Figure~\ref{fig1S2S}(b) also shows the response curves as predicted by
the simplified equations obtained from the $G\to\infty$ limit [i.e., by
collapsing all layers, $P_t^g(x)=P_t(x)$, $\forall g$]. Since they
nearly coincide with the equations for $G=10$ (which have a much
higher dimensionality), it suffices to work with $G\to\infty$, which
lends itself to analytical calculations. By expanding (around $F\simeq
0$) the single equation resulting from
Eqs.~\ref{P(2)},~\ref{P(0)},~\ref{P1g} and~\ref{Lambda} in their
stationary states, one obtains the value of critical value of
$p_\lambda$ as predicted by the 1S approximation for general $k$,
$p_\delta$ and $\beta$:
\begin{equation}
\label{eq:plambdac1S}
p^{(1S)}_{\lambda c} =   \frac{p_\delta}{k+\beta}\; .
\end{equation}
Still with $p_h = 0$ [i.e., in the absence of stimulus ($h=0$)], and recalling that $\Delta t = 1$~ms (i.e., rates $F$ and $h$ are expressed in kHz), the 1S approximation yields
the following behavior near criticality (i.e., for
$p_\lambda \gtrsim p^{(1S)}_{\lambda c}$):
\begin{equation}
\label{eq:beta}
F(h=0, \epsilon) \simeq \frac{p_\delta}{C}   {\epsilon}^{\tilde\beta}\; ,
\end{equation} 
where $\tilde\beta=1$ is a critical exponent, 
$\epsilon= \frac{p_\lambda-p^{(1S)}_{\lambda c}}{ p^{(1S)}_{\lambda c}}$, and 
 $C =\frac{k p_\delta^2}{(k+\beta)^2} \left[ \frac{(k-1)}{2} + \beta \right]
+  \frac{p_\gamma + p_\delta}{p_\gamma}$. Since in this case
the order parameter corresponds to a density of activations and the
system has no symmetry or conserved quantities, $\tilde\beta$
corresponds to the mean-field exponent of systems belonging to the
directed percolation (DP) universality class~\citep{Marro99}. 

The response function can also be obtained analytically for weak
stimuli (for $h\ll\epsilon$, $p_h \ll 1$, thus $p_h \simeq h \Delta t = h$). Below criticality ($p_\lambda <
p^{(1S)}_{\lambda c}$), the response is linear:
\begin{equation}
\label{eq:linearresponse}
F(h,\epsilon) \simeq \frac{h}{ p_\delta |\epsilon|}\; .
\end{equation}
As is usual in these cases, the linear response approximation breaks
down at $p_\lambda = p^{(1S)}_{\lambda c}$~\citep{Furtado06}. For
$p_\lambda = p^{(1S)}_{\lambda c}$, one obtains instead
\begin{equation}
\label{eq:criticalresponse}
F(h,\epsilon=0) \simeq \left(\frac{h}{C} \right)^{1/\delta_h}\; ,
\end{equation} 
where $\delta_h = 2$ is again a mean-field exponent corresponding to the
response at criticality~\citep{Marro99}.

In Fig.~\ref{fig1S2S}(c) we show in the plane $(p_\lambda, \beta)$ the
critical line given by Eq.~\ref{eq:plambdac1S}, as well as the line
obtained by numerically iterating Eqs.~\ref{P1g}-\ref{eq:LambdaG} for
finite $G$. It is interesting to note that the curves for $G\to\infty$
and $G=5, 10$ split when $\beta$ decreases. If one remembers that the
simulated model has no active phase, the resulting phase diagram
suggests that the 1S solution can perform well for $\beta \simeq
0$. Unfortunately, however, the limit $\beta\to 0$ corresponds to the
absence of backpropagating spikes, which in several cases of interest
is far from a realistic assumption (backpropagation of action
potentials well into the dendritic tree has been observed
experimentally~\citep{JohnstonReviewDendrites96, Stuart94}).

\subsection{Two-site mean-field approximation}

\begin{figure}[!ht]
\begin{center}
\includegraphics[angle=0,width=0.44\columnwidth]{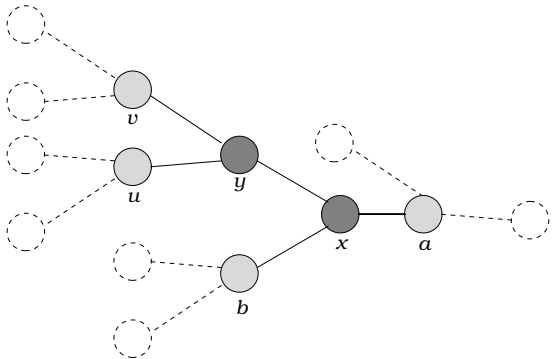}
\caption{\label{fig:Pairs} Two-site mean-field approximation schematic
  representation of a general pair ($x$ and $y$) in a binary tree
  ($k=2$).  In order to describe the dynamics of $x$ and $y$, each of
  their neighbors must be taken into account. According to
  Eq.~\ref{eq:2S}, the joint probability of the labeled sites is
  rewritten in terms of two-site probabilities. }
\end{center}
\end{figure}

The next natural step would be to consider the so-called pair or
two-site (2S) mean-field approximation~\citep{Marro99}, in which only
nearest-neighbor correlations are kept:
$P_t(x|y,u,v)\stackrel{(2S)}\approx P_t(x|y)$. In that case, the
dynamics of one-site probabilities end up depending also on two-site
probabilities~\citep{Furtado06}. Those, on their turn, depend on
higher-order terms, but under the 2S truncation these can be
approximately written in terms of one-site and two-site probabilities. The
schematic representation of a general pair of neighbor sites ($x$ and
$y$), along with their corresponding neighbors ($a$, $b$ and $u$,
$v$), is depicted in Fig~\ref{fig:Pairs}.  In the case of an infinite
tree, and restraining oneself to the isotropic case $\beta=1$, one can
drop the generation index $g$ and employ the isotropy assumption
$P_t(x,y)=P_t(y,x)$ to write the general joint probability in the
two-site approximation as
\begin{equation}
  \label{eq:2S}
P_t(a;x;y,b;u,v)\stackrel{(2S)}\approx 
\frac{P_t(a;x) P_t(x;b) P_t(x;y) P_t(y;u)P_t(y;v)}{ [P_t(x) P_t(y)]^2 }\; .
\end{equation}

In this simplified scenario, the collective dynamics is reduced to
that of a probability vector containing two-site probabilities (from which 
single-site probabilities can be obtained, please
refer to Appendix~\ref{2sequations}). Taking all normalizations into
account, the dimensionality of this vector can be reduced to 5. As can
be seen in Appendix~\ref{2sequations}, however, this simple refinement
in the mean-field approximation already leads to very cumbersome
equations.

As shown in Figs.~\ref{figCRandDRdef},~\ref{fig1S2S}(a), 
and~\ref{fig1S2S}(d), the gain in the quality of the approximation
falls far short of the increase in the complexity of the
calculations. In fact, the 1S and 2S approximations yield
qualitatively similar results, capturing the essential features of the
system behavior only for $p_\lambda$ smaller than some critical value
$p_{\lambda c}$. For $p_\lambda >p_{\lambda c}$, both approximations
predict a phase transition to self-sustained activity, with
$p_{\lambda c}=1/3$ for 1S and $p_{\lambda c}=1/2$ for 2S (in the case
$\beta=1=p_\delta$). These predictions are incorrect: when simulating
the model without external driving ($h=0$), in a few time steps
[${\cal O}(G)$] the system goes to an absorbing state~\citep{Marro99},
from which it cannot scape in the absence of further stimulation.

One can interpret the results of the approximations as follows. At the
1S approximation  level, a quiescent site will  typically be activated
by any of  its three spiking neighbors at  the phase transition, hence
$p_{\lambda c}=1/3$.  The refinement of the  2S approximation consists
in keeping track of the  excitable wave propagation from one neighbor,
leaving  two  other neighbors  (wrongly  assumed  to be  uncorrelated)
available for activity propagation, hence $p_{\lambda c}=1/2$.

One could, in principle, attempt to solve this problem by increasing the
order of the cluster approximation (keeping, e.g., 3- and 4-site
terms). However, the resulting equations are so complicated that their
usefulness would be disputable, especially for applications in
Neuroscience. It is unclear how more sophisticated mean-field
approaches (such as, e.g., non-equilibrium cavity methods~\citep{Mezard87,
  Skantzos05, Hatchett08}) would perform in this system. In principle,
they seem particularly appealing to deal with the case $p_\delta< 1$,
when a phase transition to an active state is allowed to occur (and
whose universality class is expected to coincide with that of the
contact process on trees~\citep{Pemantle92, Morrow94}). Attempts in
this direction are promising and would be welcome.

In the following section, we propose an alternative approximation
scheme which circumvents the difficulties of the regime $p_\lambda
\lesssim 1$ and at the same time takes into account finite-size
effects.


\subsection{\label{EW}Excitable-wave mean-field approximation}

The difficulties of the 1S and 2S approximations with the
strong-coupling regime are not surprising. Note that the limit of
deterministic propagation (approached in our model as $p_\lambda \to
1$) of deterministic excitations ($p_\delta=1$) is hardly handled by
continuous-time Markov processes on the lattice. To the best of our
knowledge, a successful attempt to analytically determine the scaling
of the response function to a Poisson stimulus of a hypercubic
deterministic excitable lattice was published only
recently~\citep{Ohta05} (and later confirmed in biophysically more
detailed models~\citep{Ribeiro08a}). While these scaling arguments
have not yet been adapted to the Cayley tree, the collective response
resulting from the interplay between the propagation and annihilation of
quasi-deterministic excitable waves remains an open and important
problem. In the following, we restricy ourselves to the case $p_\delta=1$, 
i.e., deterministic spike duration.  

As discussed above, the 1S and 2S approximations give poor results
essentially because they fail to keep track of where the activity
reaching a given site comes from. We therefore propose here an
excitable-wave (EW) mean-field approximation which attempts to address
precisely this point. 

\begin{figure}[!ht]
\begin{center}
\includegraphics[angle=0,width=0.57\columnwidth]{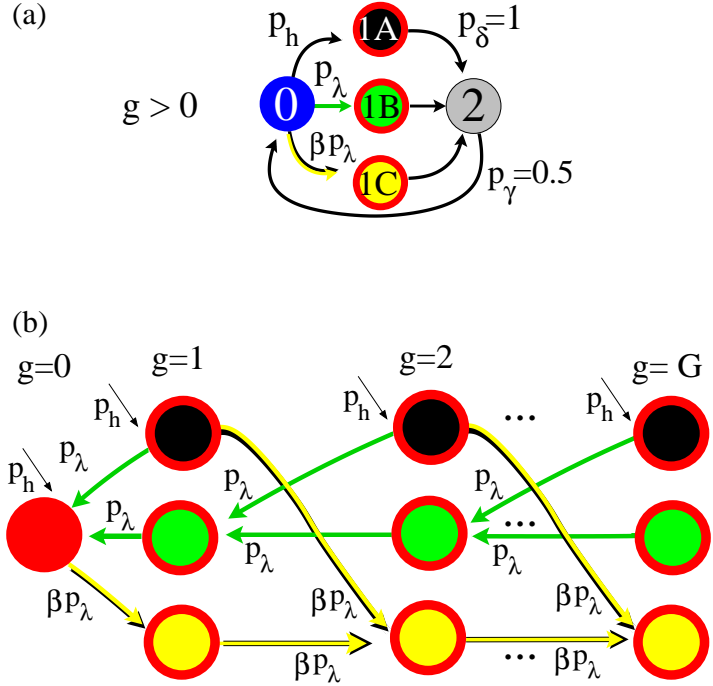}
\caption{\label{fig:meanfield}(Color online) Schematic representation
  of the excitable-wave mean-field approximation.  (a) Dynamics of
  each layer ($g>0$) in the excitable-wave (EW) mean-field
  approximation (see text for details).  There are 3 different active
  states: 1A represents activity coming from an external input; 1B is
  reached due to forward activity from the next layer, whereas 1C is
  excited by backpropagating activity from the previous layer. (b)
  Schematic EW mean-field approximation dynamics in a tree with $G$
  layers. Note that there are no loops in the activity flux.}
\end{center}
\end{figure}

The rationale is simple: in an excitable tree, activity can always be
decomposed in forward- and backward-propagating excitable
waves. Formally, this is implemented as follows. We separate (for
$g>0$) the active state (1) into three different active states: 1A,
1B, and 1C, as represented in Fig.~\ref{fig:meanfield}(a). $P_t^g(1A)$
stands for the probability that activation (at layer $g$ and time $t$)
was due to the input received from an external source (controlled by
$p_h$). The density of elements in 1A can excite quiescent neighbors
at both the previous and the next layers. $P_t^g(1B)$ corresponds to
the density of elements in layer $g$ which were quiescent at time $t-1$
and received input from the next layer ($g+1$) (i.e., a forward
propagation).  The density of elements in 1B can excite solely
quiescent neighbors at the previous layer.  Finally, $P_t^g(1C)$
accounts for the activity coming from the previous layer (i.e.,
backpropagation).  The density of elements in 1C can excite solely
quiescent neighbors at the next layer. For lack of a better name, we
refer to these different virtual states as excitation
components. Figure~\ref{fig:meanfield}(b) represents the activity flux
in the dendritic tree as projected by the EW mean-field
approximation. The absence of loops guarantees the suppression of the
spurious non-equilibrium phase transition predicted by the traditional
cluster expansions.

Following these ideas, one can write the equations for the $g>0$ layers as
\begin{eqnarray}
\label{P1A}P^g_{t+1}(1A)&=& P^g_t(0) \Lambda^g_A  \; , \\ 
\label{P1B}P^g_{t+1}(1B)&=& P^g_t(0) (1-\Lambda^g_A) \Lambda^g_B(t)   \; , \\ 
\label{P1C}P^g_{t+1}(1C)&=& P^g_t(0) (1-\Lambda^g_A)\left[1-\Lambda^g_B(t)\right]\Lambda^g_C(t)  \; ,
\end{eqnarray} 
where, in analogy with Eq.~\ref{Lambda}, the excitation probabilities
are now given by
\begin{eqnarray}
\label{LambdaA}\Lambda^g_A &=& p_h,  \\
\label{LambdaB}\Lambda^g_{B} (t)&=& 1-\left\lbrace 1-p_\lambda \left[P^{g+1}_t(1A)+P^{g+1}_t(1B)\right] \right\rbrace^k, \\
\label{LambdaC}
\Lambda^g_{C} (t)&=& \beta p_\lambda \left[P^{g-1}_t(1A)+P^{g-1}_t(1C)\right].
\end{eqnarray}
Equations~\ref{P(2)}~and~\ref{P(0)} remain unchanged, with $
P^{g}_{t}(1) \equiv P^g_{t}(1A)+P^g_{t}(1B)+P^g_{t}(1C)$. The dynamics
of the most distal layer $g=G$ is obtained by fixing
$\Lambda_B^{G}(t)=0$. The apical ($g=0$) element has a simpler
dynamics since it does not receive backpropagating waves, so its
activity is governed by Eq.~\ref{P1g}, with $g=0$ and $\Lambda^0(t) =
1-(1-p_h)\left\lbrace 1-p_\lambda
  \left[P^{1}_t(1A)+P^{1}_t(1B)\right]\right\rbrace^{k+1}$ instead of
Eq.~\ref{Lambda}. Taking into account the normalization conditions,
the dimensionality of the map resulting from the EW approximation is
$4(G-1)+5$. 

It is important to notice that, while Eqs.~\ref{LambdaA}-\ref{LambdaC}
are relatively straightforward, there is a degree of arbitrariness in
the choice of Eqs.~\ref{P1A}-\ref{P1C}. As written, they prescribe an
\textit{ad hoc} priority order for the recruitment of the excitation components
of the EW equations: first by synaptic stimuli (Eq.~\ref{P1A}), then
by forward propagating waves (Eq.~\ref{P1B}), and finally by
backpropagating waves (Eq.~\ref{P1C}). This choice seems to be
appropriate in the regime of weak external driving, insofar as the
order coincides with that of the events observed in the experiments:
forward dendritic spikes, a somatic spike, then backpropagating
dendritic spikes~\citep{Stuart94}. Appendix~\ref{priorityOrder}
compares the response functions for different priority orders to
emphasize the robustness of the approximation with respect to that.

\begin{figure}[!ht]
\centerline{\includegraphics[angle=0,width=0.99\columnwidth]{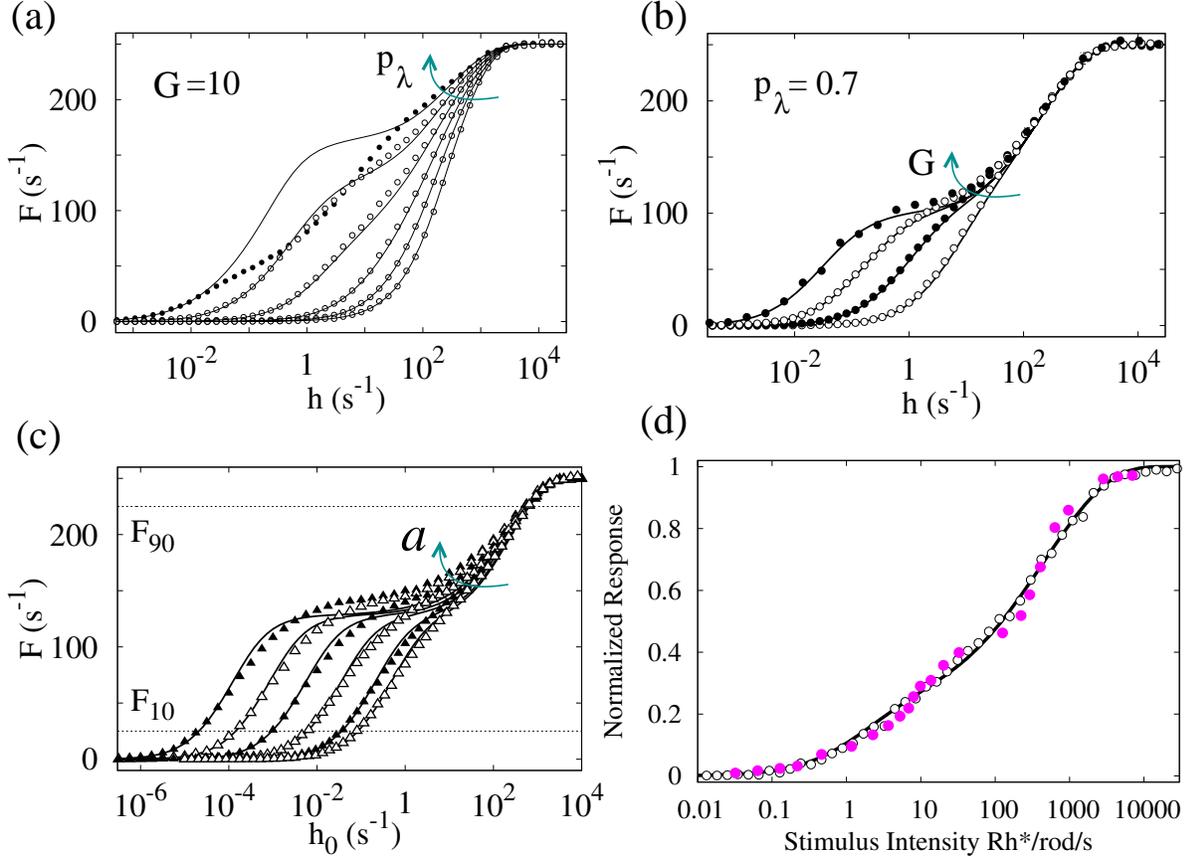}}
\caption{\label{figCR}(Color online) Response functions: simulations
  compared to the EW approximation (both with $\beta=1$) and
  experimental data. (a) Family of response functions for $G=10$ and
  $p_\lambda=0, 0.2,0.4,\ldots, 1$. Symbols are the simulations (as in
  Ref.~\cite{Gollo09}) and solid curves are the EW mean-field
  approximation. (b) Family of response functions for $p_\lambda=0.7$
  and different tree sizes: $G=5, 10, 15, 20$.  (c) Simulations and EW
  response function for external stimuli spatially distributed as
  $h(g) =h_0 e^{a g}$: from right to left, $a=0, 0.1, 0.3, \ldots,
  0.9$ ($p_\lambda=0.8$ and $G=10$). Horizontal lines are plotted for
  the estimation of the dynamic range. (d) Experimental result from mouse
  retinal ganglion cells shows double sigmoid response curves
  (closed symbols) as a function of stimulus $I$ (measured in
  rhodopsins/isomerizations/rod/second~\citep{Deans02}). They can be
  reasonably well fit by both simulations (open symbols,
  $p_\lambda=0.58$ and $h=0.37$ I) and the EW response function
  ($p_\lambda=0.59$ and $h=0.4$ I) with tree size $G=15$.}
\end{figure}

Though noncontrolled, the EW mean-field approximation does provide
excellent agreement with simulations. The results for $G=10$ can be seen
in Fig.~\ref{figCR}(a), which shows a family of response curves $F(h)$
for varying coupling $p_{\lambda}$. One observes that the EW
mean-field results (lines) follow the simulation results (symbols)
very closely up to $p_\lambda \simeq 0.8$, reproducing even the
double-sigmoidal behavior of the curves~\citep{Deans02, Gollo09,
  Ferrante09, Jedlicka10}. For larger values of $p_\lambda$, agreement
is restricted to very small or very large values of $h$ (for
intermediate values of $h$, note that $F(p_\lambda)$ is nonmonotonous,
a rather counterintuitive phenomenon called ``screening
resonance''~\citep{Gollo09}). Most importantly, however, the EW
equations eliminate the phase transition wrongly predicted by the
traditional mean-field approximations.


In a real neuron, the number of layers is finite [${\cal O}(10)$ or
less] and it would be extremely interesting to have an analytical
approximation which managed to take finite-size effects into
account. As it turns out, the mean-field approximation we propose can
do precisely that, since it couples densities at different layers (so
$G$ again controls the dimensionality of the mean-field
map). Figure~\ref{figCR}(b) compares simulations (symbols) with the
stationary state of the EW mean-field equations (lines) for different
system sizes. Note that the agreement is excellent from $G=5$ up to
$G=20$, for the whole range of $h$ values.

The EW mean-field approximation is also very robust against previously
proposed variants of the model. For instance, in several neurons the
distribution of synaptic inputs along the dendritic tree is
nonuniform, increasing with the distance from the soma. A
one-parameter variant which incorporates this nonuniformity consists
in a layer-dependent rate $h(g)=h_0 e^{a g}$~\citep{Gollo09}, as
described in section~\ref{model}. Figure~\ref{figCR}(c)
depicts a good agreement between simulations and the EW approximation
for a range of $a$ values.

Also shown in Fig.~\ref{figCR}(d) is a comparison among experimental
results from retinal ganglion cells to varying light
intensity~\citep{Deans02} (closed symbols), simulations (open symbols), 
and EW mean-field approximation (lines), which agree reasonably
well. Therefore the approximation we propose can, in principle, be
useful for fitting experimental data and reverse-engineering parameter
values from data-based response functions at a relatively small
computational cost (say, compared to simulations). In this particular
example, it is important to emphasize that the experimental response
curves are, in principle, influenced by other retinal elements. Given
the very simple nature of our model, it is hard to pinpoint which part
of the retinal circuit our Cayley tree would represent. Following
Shepherd~\cite{Shepherd,Gollo09}, however, we suggest that the
ganglionar dendritic arbor plus the retinal cells connected to it by
gap junctions (electrical synapses) can be viewed as an extended
active tree similar to the one studied here, with a large effective
$G$.

\begin{figure}[!hb]
\centerline{\includegraphics[angle=0,width=0.85\columnwidth]{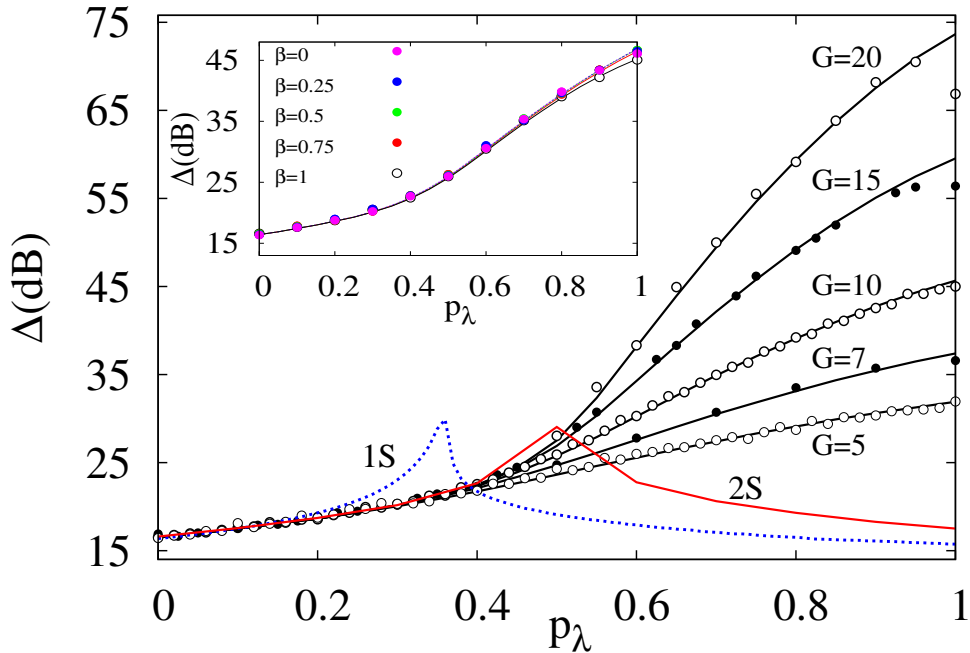}}
\caption{\label{fig:DR}(Color online) Dynamic range as a function of
  the coupling parameter for 1S ($G=10$), 2S 
  (infinite tree), EW (black lines) mean-field approximations compared
  to simulations (symbols, as in Ref.~\cite{Gollo09}) for different
  trees sizes $G$. The inset shows the dynamic range of dendritic
  trees subjected to an asymmetrical activity propagation probability
  controlled by $\beta$ (symbols stand for the simulations and the
  curves for the EW approximation for $G=10$).}
\end{figure}

The dynamic range $\Delta$ is one of the features of the response
function which has received attention in the literature in recent
years~\citep{Copelli02,Copelli05a,Copelli05b,Furtado06, Kinouchi06a,
  Copelli07, Wu07, Assis08, Ribeiro08a, Publio09, Shew09,
  Larremore11, Larremore11b, Buckley11}.  Here it serves the purpose of summarizing the quality
of the EW mean-field approximation in comparison with model
simulations. In Fig.~\ref{fig:DR} we plot $\Delta$ as a function of
$p_\lambda$ for several system sizes $G$. Both the 1S and 2S
mean-field approximations predict a non-equilibrium phase transition in
the model, where a peak of the dynamic range therefore
occurs~\citep{Kinouchi06a}. Both approximations perform badly especially
in the high-coupling regime. The EW approximation correctly predicts
the overall behavior of the $\Delta(p_\lambda)$ curves, for all system
sizes we have been able to simulate. Finally, the inset of
Fig.~\ref{fig:DR} shows a second variant of the model in which the
parameter $\beta$, which controls the probability of a spike
backpropagating, is free to change. Once more, the EW approximation
manages to reproduce the $\Delta(p_\lambda)$ curves obtained from
simulations for the full range of $\beta$ values.

\section{Conclusions}
\label{conclusions}

The need for a theoretical framework to deal with active dendrites has
been largely recognized. However, the plethora of physiological
details which are usually taken into account to explain local
phenomena renders the problem of understanding the dynamics of the
tree as a whole analytically untreatable. What we have proposed is the
use of a minimalist statistical physics model in which our ignorance
about several physiological parameters is thrown into a single
parameter $p_\lambda$. The model has provided several insights and
predictions, most of them yet to be tested
experimentally~\citep{Gollo09}. Here we have shown that the model is
amenable to analytical treatment as well.

We have compared different mean-field solutions to the model, and
shown that standard cluster approximations (1S and 2S) yield poor
results. They incorrectly predict phase transitions which are not
allowed in the model, thereby failing to reproduce the response
functions precisely in the low-stimulus and highly nonlinear regime
(where most of the controversies are bound to
arise~\citep{Bhandawat07}).

To overcome this scenario we developed an excitable-wave mean-field
approximation which takes into account the direction in which the
activity is propagating through the different layers of the
tree. Though \textit{ad hoc}, the approximation reproduces simulation results
with very reasonable accuracy, for a wide range of parameters and two
biologically relevant variants of the model. We hope that our EW
mean-field approximation may therefore contribute to the theoretical
foundation of dendritic computation~\citep{Abbott08}.

It is important to recall that the theory attempts to address a model
in a regime which is expected to be close to that of a neuron \textit{in vivo}:
dendritic spikes are generated at random, may or may not propagate
along the dendrites, and annihilate each other upon collision. Our
model allows one to formulate theoretical predictions for \textit{in vivo}
experiments in sensory system neurons, e.g., ganglion
cells~\citep{Deans02}, olfactory mitral cells~\citep{Wachowiak01} or
their insect counterparts, i.e., antennal lobe projection
neurons~\citep{Bhandawat07}: a) if one removes part of the dendritic
tree and/or b) blocks the ionic channels responsible for dendritic
excitability, one should see a decrease in the neuronal dynamic
range. To the best of our knowledge, these experiments have not been
done yet.

Another issue upon which our results could have a bearing is the so-called
gain control modulation. In the neuroscience literature, the term
refers to the neuronal capacity to change the slope of the
input-output response function~\citep{Rothman09}. This property has
been reported in visual~\citep{Tovee94, Brotchie95,Treue99,
  Anderson00}, somatosensory~\citep{Yakusheva07} and
auditory~\citep{Ingham05} systems. Several possible mechanisms have
already been proposed to explain gain control, based on synaptic
depression~\citep{Abbott97, Rothman09}, background synaptic
input~\citep{Chance02, Fellous03}, noise~\citep{Hansel02}, shunting
inhibition~\citep{Mitchell03, Prescott03} and excitatory
(NMDA~\citep{Berends05}) as well as inhibitory
(GABA$_A$~\citep{Semyanov04}) ionotropic receptor dynamics.

All these mechanisms are intercellular, in the sense that they rely on
the influence of factors external to the neuron. Our model, on the
other hand, shows gain control in its dependence on the coupling
parameter $p_\lambda$, which controls the propagation of dendritic
spikes {\em within\/} the tree. It is therefore an intracellular
mechanism which offers an additional explanation for this ubiquitous
phenomenon. 

The physics of complex systems is becoming more and more embracing,
shedding light in different areas, including neuroscience.
Particularly at the cellular and subcellular levels, we foresee the
merging of the two fields, dendritic computation and statistical
physics, as a promising avenue.  The maturity of the latter could
illuminate several frontiers in neuroscience.

\section{Acknowledgments}
The authors acknowledge financial support from the Brazilian agency
CNPq. LLG and MC have also been supported by FACEPE, CAPES and special
programs PRONEX. MC has received support from PRONEM and INCeMaq. This
research was also supported by a grant from the MEC (Spain) and
FEDER under project FIS2007-60327 (FISICOS) to LLG. LLG is
appreciative to Prof. John Rinzel and his working group for the
hospitality and valuable discussions during his 3-month visit,
March to May, 2009 in the Center for Neural Science at New York
University.
\\


\appendix

\section{Derivation of the general master equation}
\label{derivation}

Let us illustrate how the master equation is obtained by starting with
the simplest possible case, namely, the latest layer of the Cayley
tree. Sites at the surface connect to a single site (their mother
branchlet), so the probability of their being excited at time $t+1$ is 
\begin{equation}
  \label{eq:PG}
  P_{t+1}^G(;1;) = [1-(1-p_h)(1-\beta p_\lambda)]P_t^G(1;0;)
  +p_h\sum_{l\neq 1}P_t^G(l;0;)+(1-p_\delta)P_t^G(;1;)\;.
\end{equation}
Each term has a straightforward interpretation: the first term
corresponds to the probability that the surface site is quiescent
(state 0), its neighbor is active (state 1), and excitation gets to
the surface via an external stimulus ($p_h$) and/or backpropagating
transmission ($\beta p_\lambda$); the second term corresponds to the
excitation of the surface site via an external stimulus, provided that
it is quiescent (0) and its neighbor is in any state other than active
(1); the third term corresponds to the probability that the surface
site was in state 1 and did not move to state 2 (a transition
controlled by $p_\delta$, see Fig.~\ref{fig:dendrite}c).

The next easiest case to consider is that of the root ($g=0$) site. It
connects to $k+1$ daughter branchlets, and can be excited by any
number of them. Contrary to the $g=G$ surface sites, the root site
only receives forward propagating activity (hence $\beta$ plays no
role). Analogously to Eq.~\ref{eq:PG}, the equation for $P_{t+1}^0$ is
given by
\begin{eqnarray}
\label{zero}
P_{t+1}^{\,0}(;1;) & = & \left[1-(1-p_h)(1-p_\lambda)^{k+1}\right] P_t^{\,0}(;0;1^{(k+1)})  {k+1 \choose k+1} \nonumber \\ 
 & &+ \left[1-(1-p_h)(1-p_\lambda)^{k}\right]\sum_{j_1 \ne 1}  P_t^{\,0}(;0;1^{(k)},j_1 ) {k+1 \choose k} \nonumber \\ 
 & &+ \left[1-(1-p_h)(1-p_\lambda)^{k-1}\right]\sum_{j_1,j_2 \ne 1}  P_t^{\,0}(;0;1^{(k-1)},j_1, j_2) {k+1 \choose k-1} \nonumber \\ 
 & &  + \;\cdots\; \nonumber \\
 & &  + 
 \left[1-(1-p_h)(1-p_\lambda)^{m}\right]\sum_{j_1,j_2,\ldots,j_{k+1-m}
   \ne 1}  
P_t^{\,0}(;0;1^{(m)},j_1, j_2,\ldots,j_{k+1-m}) {k+1 \choose m} \nonumber \\
 & &  + \;\cdots\; \nonumber \\
 & &+ \left[1-(1-p_h)(1-p_\lambda)\right]\sum_{j_1,\ldots,j_k \ne 1}  P_t^{\,0}(;0;1,j_1,\ldots,j_k) {k+1 \choose 1} \nonumber \\ 
 & &+ p_h \sum_{j_1,\ldots,j_{k+1} \ne 1} P_t^{\,0}(;0;j_1,\ldots,j_{k+1}) \nonumber \\
 & &+ (1-p_\delta) P_t^{\,0}(;1;) \;.
\end{eqnarray}
The terms of the kind $ \left[1-(1-p_h)(1-p_\lambda)^{m}\right]$
account for the excitation of the root site via an external stimulus
and/or via transmission from $m$ of its active daughter branchlets,
regardless of the state of its $k+1-m$ non-active neighbors (hence the
sum over $j_1,j_2,\ldots,j_{k+1-m} \ne 1$). Each term is weighted by
the number ${k+1 \choose m}$ of combinations of $m$ active sites out
of $k+1$. The sum of these terms (from $m=1$ to $m=k+1$) therefore
plays a role equivalent to that of the first term in
eq.~\ref{eq:PG}. The two last terms are analogous to those of
eq.~\ref{eq:PG}. 

Finally, we come to the equation for a general site with $1\leq g \leq
G-1$, which can be excited by both its mother as well as its daughter
branchlets. The equation for $P_{t+1}^g$ thus generalizes the terms of
the preceding equations:
\begin{eqnarray}
\label{eq:Pg}
P_{t+1}^{g}(;1;) & = & \left[1-(1-p_h)(1-\beta p_\lambda)(1-p_\lambda)^{k} \right] P_t^{g}(1;0;1^{(k)})       
{k \choose k} \nonumber \\
 & &+\left[1-(1-p_h)(1-\beta p_\lambda)(1-p_\lambda)^{k-1} \right] \sum_{j_1 \ne 1}P_t^{g}(1;0;1^{(k-1)},j_1)
{k \choose k-1} \nonumber \\
 & & + \;\cdots\; \nonumber \\
 & & +\left[1-(1-p_h)(1-\beta p_\lambda)(1-p_\lambda)\right]\sum_{j_1,\ldots,j_{k-1} \ne 1} P_t^{g}(1;0;1,j_1,\ldots,j_{k-1})
{k \choose 1} \nonumber \\
 & & + \left[1-(1-p_h)(1-\beta p_\lambda)\right] \sum_{j_1,\ldots,j_k \ne 1} P_t^{g}(1;0;j_1,\ldots,j_k)
{k \choose 0} \nonumber \\
 & &+ \left[1-(1-p_h)(1-p_\lambda)^{k}\right] \sum_{\ell \ne 1} P_t^{g}(\ell;0;1^{(k)})
{k \choose k} \nonumber \\
 & &+ \left[1-(1-p_h)(1-p_\lambda)^{k-1}\right] \sum_{\ell,j_1 \ne 1} P_t^{g}(\ell;0;1^{(k-1)},j_1)
{k \choose k-1} \nonumber \\
 & & +\;\cdots\; \nonumber \\
 & & +\left[1-(1-p_h)(1-p_\lambda)\right]\sum_{\ell,j_1,\ldots,j_{k-1} \ne 1} P_t^{g}(\ell;0;1,j_1,\ldots,j_{k-1})
{k \choose 1} \nonumber \\
 & &+ p_h\sum_{\ell,j_1,\ldots,j_{k} \ne 1} P_t^{g}(\ell;0;j_1,\ldots,j_{k}){k \choose 0} \nonumber \\
 & &+(1-p_\delta ) P_t^{g}(;1;)\;.
\end{eqnarray}

Equations~\ref{eq:PG}-\ref{eq:Pg} can be drastically
simplified~\citep{Furtado06}. Taking into account the normalization
condition
 \begin{equation}
 P_{t}^{g}(a;b;j_1,\ldots,j_{\ell-1})\equiv \sum_{j_{\ell}}P_{t}^{g}(a;b;j_1,\ldots,j_{\ell-1},j_{\ell})\;,
\end{equation}
the sums in eqs.~\ref{eq:PG}-\ref{eq:Pg} can be reduced. For instance, 
\begin{eqnarray}
\label{somatorio}
\sum_{j_1 \ne 1}  P_t^{g}(\ell;0;1^{(k-1)},j_1) & = & \sum_{j_1}P_t^{g}(\ell;0;1^{(k-1)},j_1) - P_t^{g}(\ell;0;1^{(k)})\nonumber \\
& = & P_t^{g}(\ell;0;1^{(k-1)}) - P_t^{g}(\ell;0;1^{(k)})\;.
\end{eqnarray}
Iterating this procedure and rearranging terms, one finally arrives at
\begin{eqnarray}
\label{0final}
P_{t+1}^0(;1;) & = & P_{t}^0(;0;1^{(0)}) - \left(1 -p_h\right)\left[  \sum^{k+1}_{i=0} p^i_{\lambda} {k+1 \choose i} (-1)^i P_t^0(;0;1^{(i)})  \right] \nonumber \\
 & & +(1-p_ \delta) P_t^{0}(;1;)\;,
\end{eqnarray}

\begin{eqnarray}
\label{etafinal}
P_{t+1}^{g}(;1;) & = & P_{t}^{g}(;0;1^{(0)}) - \left(1- p_h \right) \Biggl[  \sum^{k}_{i=0}\biggl(  p^i_{\lambda} {k \choose i} (-1)^i P_t^{g}(;0;1^{(i)}) \nonumber \\ 
 & &- \beta p^{i+1}_{\lambda}{k \choose i}(-1)^i P_t^{g}(1;0;1^{(i)})\biggr)\Biggr]+(1-p_ \delta) P_t^{g}(;1;)\;, 
\end{eqnarray}

\begin{eqnarray}
\label{Gfinal}
P_{t+1}^G(;1;) & = & P_{t}^G(;0;1^{(0)}) - \left(1- p_h \right)\left( P_t^G(;0;1^{(0)})-\beta p_{\lambda}P_t^G(1;0;1^{(0)})  \right)  \nonumber \\
 & & +(1-p_ \delta) P_t^{G}(;1;)\;.
\end{eqnarray}
Recalling that $P_t^g(x;y;w^{(0)})\equiv P_t^g(x;y;)$, we recover
eqs.~\ref{P(1)},~\ref{P0(1)} and~\ref{PG(1)}. 

\section{Two-site mean-field equations}
\label{2sequations}

For an infinite dendritic tree, the complete set of equations under the 2S 
approximation is given by:

\begin{eqnarray}
\label{finalP00}
P_{t+1}(0;0) &\stackrel{(2S)}\approx& \{ (1-p_h)^2 P_t(0;0) [p_{\lambda}P_t(0;1)- P_t(0)]^4    \nonumber \\
 & & +2 p_{\gamma} (1-p_h) P_t(0;2) P_t(0) ^2  [p_{\lambda}P_t(0;1)- P_t(0)]^2   \nonumber \\
 & & + p_{\gamma} ^2 P_t(2;2) P_t(0) ^4 \} \frac{1}{P_t(0) ^2}\;;
\end{eqnarray}

\begin{eqnarray}
\label{finalP01}
P_{t+1}(0;1) &\stackrel{(2S)}\approx& -\Bigl( P_t(0;1) P_t(0)^2 \{ -(1-p_{\delta})(1-p_h)(1-p_{\lambda}) [p_{\lambda}P_t(0;1)- P_t(0)]^2\}   \nonumber \\
 & &   + (1-p_h) P_t(0;0) \{ (1-p_h) [p_{\lambda}P_t(0;1)- P_t(0)]^4  \nonumber \\
 & &  - P_t(0)^2 [p_{\lambda}P_t(0;1)- P_t(0)]^2\} \nonumber \\
 & &  +  p_{\gamma}  P_t(0;2)\{  (1-p_h) P_t(0)^2  [p_{\lambda}P_t(0;1)- P_t(0)]^2 - P_t(0)^4 \}  \nonumber \\ 
 & &  - (1- p_{\delta}) p_{\gamma}   P_t(1;2)  P_t(0)^4  \Bigr) \frac{1}{P_t(0)^4}\;;
\end{eqnarray}

\begin{eqnarray}
\label{finalP02}
P_{t+1}(0;2) &\stackrel{(2S)}\approx& \{ (1-p_{\gamma}) (1-p_h) P_t(0;2) [p_{\lambda}P_t(0;1)- P_t(0)]^2  \nonumber \\
 & & + p_{\delta}  (1-p_h) (1-p_{\lambda}) P_t(0;1)  [p_{\lambda}P_t(0;1)- P_t(0)]^2  \nonumber \\
 & & + p_{\gamma} (1-p_{\gamma}) P_t(2;2) P_t(0)^2  \nonumber \\
 & & + p_{\delta} p_{\gamma}  P_t(1;2) P_t(0)^2  \} \frac{1}{P_t(0)^2}\;;
\end{eqnarray}

\begin{eqnarray}
\label{finalP11}
P_{t+1}(1;1) &\stackrel{(2S)}\approx& \Bigl( (1-p_h)  P_t(0;0)\{(1-p_h)[p_{\lambda}P_t(0;1)- P_t(0)]^4  \nonumber \\ 
 & &   -2 P_t(0)^2 [p_{\lambda}P_t(0;1)- P_t(0)]^2 +P_t(0)^4\}   \nonumber \\
 & &   -2 P_t(0;1) P_t(0)^2 \{ (1-p_{\delta})(1-p_h)(1-p_{\lambda}) [p_{\lambda}P_t(0;1)- P_t(0)]^2  \nonumber \\
 & & -(1-p_{\delta})P_t(0)^2\} +  (1-p_{\delta})^2 P_t(1;1)  P_t(0)^4 \Bigr) \frac{1}{P_t(0)^4}\;;
\end{eqnarray}

\begin{eqnarray}
\label{finalP12}
P_{t+1}(1;2) &\stackrel{(2S)}\approx& - \Bigl( P_t(0;2) \{  (1-p_{\gamma}) (1-p_h)[p_{\lambda}P_t(0;1)- P_t(0)]^2 - (1-p_{\gamma}) P_t(0)^2 \} \nonumber \\
 & & +  p_{\delta} P_t(0;1) \{ (1-p_h) (1-p_{\lambda}) [p_{\lambda}P_t(0;1)- P_t(0)]^2 - P_t(0)^2 \} \nonumber \\
 & & -p_{\delta} (1-p_{\delta}) P_t(1;1) P_t(0)^2    \nonumber \\
 & & - (1-p_{\delta}) (1-p_{\gamma}) P_t(1;2) P_t(0)^2  \Bigr) \frac{1}{P_t(0)^2}\;;
\end{eqnarray}

\begin{eqnarray}
\label{finalP22}
P_{t+1}(2;2) &\stackrel{(2S)}\approx& p_{\delta}^2 P_t(1;1)+2 p_{\delta}  (1-p_{\gamma}) P_t(1;2)+ (1-p_{\gamma})^2 P_t(2;2)\;.
\end{eqnarray}
In order to solve it numerically, by normalization we make use of:
\begin{eqnarray}
\label{final2SP0}
P_{t}(0) &=& P_{t}(0;0)+ P_{t}(0;1)+ P_{t}(0;2)\;.
\end{eqnarray}
To compare the results with the simulations and experiments 
we use $F \equiv  \frac{1}{\Delta t} \lim_{t\to\infty} P_t(1)$, where: 
\begin{eqnarray}
\label{final2SF}
P_{t}(1) &=& P_{t}(0;1)+ P_{t}(1;1)+ P_{t}(1;2)\;.
\end{eqnarray}
Finally, by completeness, the last single-site probability can be obtained by:
\begin{eqnarray}
\label{final2SP2}
P_{t}(2) &=& P_{t}(0;2)+ P_{t}(1;2)+ P_{t}(2;2)\;.
\end{eqnarray}

\section{Robustness with respect to the priority order of the
  excitable components}
\label{priorityOrder}

\begin{figure}[!ht]
\centerline{\includegraphics[angle=0,width=1.0\columnwidth]{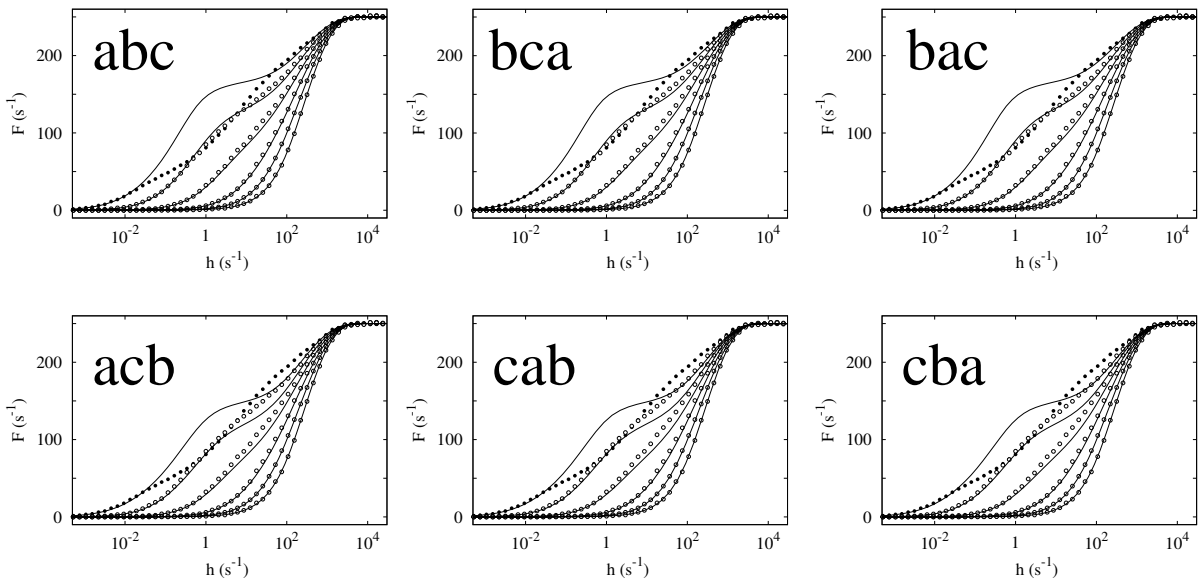}}
\caption{\label{RF} Effects of the priority order of the excitable
  components on the response functions: simulations compared to the
  EW approximation and experimental data.  Family of response
  functions for $G=10$ and $p_\lambda=0, 0.2,0.4,\ldots, 1$.  Top
  panels display combinations of 1B prior to 1C, and bottom panels the
  converse.  Symbols are the simulations and solid curves are the EW
  mean-field approximation. }
\end{figure}

The priority order used in Eqs.~\ref{P1A}-\ref{P1C} was ``ABC'', i.e.,
first 1A, followed by 1B and then 1C. The virtual state 1B accounts
for the forward-propagating excitable-wave flux whereas state 1C
accounts for the backward-propagating excitable-wave flux. In order to
compare all the different combinations, Fig.~\ref{RF} displays families
of response functions.  

Intuitively, the neuronal firing rate $F$ increases as the
forward-propagating excitable-wave flux grows.  Summarizing the
results, switching the order of 1B and 1C (lower panels in
Fig.~\ref{RF}), we reduce the forward-propagating excitable-wave flux,
and consequently, the response functions corresponding to large
$p_\lambda$ values present a lower firing rate. Moreover, the result
is virtually the same irrespective of the order in which 1A appears
(compare panels horizontally in Fig.~\ref{RF}).

The approximation is robust with respect to the order chosen, and only
minor differences can be found for strong coupling ($p_\lambda \sim
1$) and the intermediate amount of external driving input. Changes in the
order of the components modify the response functions quantitatively.
However, qualitatively, the response functions are very alike. Most of
the differences occur at the high-coupling regime ($p_\lambda \sim
1$), but do not affect the localization of $h_{10}$ and $h_{90}$
(which implies that the dynamic range remains unchanged).


\bibliography{PRE2010GolloCopelli}

\end{document}